 \documentstyle[aaspp4,graphicx]{article}
 \def\:{\mskip\medmuskip}                         
 \def\kelvin{\thinspace\rm{\sp{o}{\kern-.08333em }K}\ }
\begin{document}
\title{On the Thermonuclear Runaway in Type Ia Supernovae:}
\title{ How to run away?}
\bigskip
\bigskip
\author{P. H\"oflich}  
\affil{$^1$Department of Astronomy, University of Texas,
Austin, USA \\pah@hej1.as.utexas.edu}
\author{J. Stein}
\affil{$^2$ Racah Institute of Physics, The Hebrew
University of Jerusalem, Israel \\yossi@phys.huji.ac.il}

\begin{abstract}
 Type Ia Supernovae are thought to be thermonuclear explosions
of massive white dwarfs (WD). We present the first study of
multi-dimensional effects during the final hours prior to the
thermonuclear runaway which leads to the explosion.
 The calculations utilize an implicit, 2-D
hydrodynamical code.   Mixing  and  the ignition process are studied in detail.
 We find that the initial chemical structure of the
 WD is changed, but the material
is not fully homogenized. In particular, the exploding WD sustains a
central region with a low C/O ratio.
  This implies that the explosive nuclear burning will begin in a partially carbon-depleted
environment. The thermonuclear runaway happens  in a well-defined
region close to the center.
 It is induced by compressional heat when matter is brought inwards
by convective flows. We find no evidence for multiple spot or strong off-center
ignition. 
  Convective velocities in the WD are of the order of 100 km/sec which is well above the effective
burning speeds in SNe~Ia previously expected right after the runaway. In our calculations, the ignition
occurs near the center. Then, for $\approx$ 0.5 to 1 sec,
 the speed of the burning front  will neither be determined   
by the laminar speed nor the Rayleigh-Taylor instabilities but by convective flows
produced prior to the runaway.
  The consequences are discussed for our understanding
of the detailed physics of the flame propagation,
the deflagration to detonation transition, and the
nucleosynthesis in the central layers.
 Our results strongly suggest the     
pre-conditioning of the  progenitor as a key factor for our understanding of the
diversity in Type Ia Supernovae.
\end{abstract}
 
\section{Introduction}

 Type Ia Supernovae (SNe~Ia) are among the most spectacular events because they reach the
same brightness as an entire galaxy. This makes them good candidates to
determine extragalactic distances and to measure the basic cosmological parameters.
 Moreover, they are thought to be the major contributors to the chemical
enrichment of the interstellar matter with heavy elements.
   Energy injection by SN into the interstellar medium, triggered star formation and feedback in
galaxy formation are regarded as a key for our  understanding
of the formation and evolution of galaxies.

There is general agreement that SNe~Ia result from some  process of combustion of a
degenerate, C/O white dwarf (Hoyle \& Fowler 1960). Within this general picture, three
classes of models have been considered: (1) An explosion of a CO-WD, with mass
close to the Chandrasekhar mass, which accretes mass through Roche-lobe overflow
from an evolved companion  star (Whelan \& Iben 1973). The
explosion is then triggered by compressional heating near the WD center.  (2) An
explosion of a rotating configuration formed from the merging of  two low-mass WDs,
caused by the loss of angular momentum due to gravitational radiation
(Webbink 1984, Iben \& Tutukov 1984, Paczy\'nski 1985).  
 (3) Explosion of a low mass CO-WD
triggered by the detonation of a helium  layer (Nomoto 1980, Woosley et al. 1980, 
Woosley \& Weaver 1986).
Only the first two models appear to be viable.
The third, the sub-Chandrasekhar WD model, has been ruled out on the basis of
predicted light curves and spectra (H\"oflich et al. 1996a, Nugent et al. 1997).

 For the identification of the most common scenario, the main problem is
related to the insensitivity of the  WD structure to the progenitor star and system.
However, the last decade has witnessed an explosive growth of high-quality
data and  advances in the models for supernovae which opened up new
opportunities to constrain the physics of supernovae. For the first time, a direct connection
with the progenitors seems to be within reach. In particular, there is mounting evidence for a
connection between the properties of the progenitor, and the physics of the explosion.

 The explosion of a WD with $M_{Ch}$ is  the most likely candidate for the majority of
  `normal' SNe~Ia.
In particular,
 delayed detonation (DD) models (Khokhlov 1991, Woosley \& Weaver 1994, Yamaoka et al. 1992)
have been found to reproduce the majority of optical and infrared light curves (LC)
and spectra of SNe~Ia reasonably well
 (H\"oflich 1995; H\"oflich \& Khokhlov 1996; 
 Fisher et al. 1998; Nugent et al. 1997; Wheeler et al.  1998; Lentz et al. 2000; 
Gerardy et al. 2001).
We note that detailed analyzes of observed spectra and light curves indicate that
mergers and deflagration models such as W7 may contribute to the supernovae population
(Harkness 1987, H\"oflich \& Khokhlov 1996, Hatano et al. 2000).
  The evidence against pure deflagration models for the majority of SNeIa includes
 IR-spectra which show signs of explosive carbon burning at high 
expansion velocities (e.g. Wheeler et al. 1998) and recent calculations for 3-D deflagration fronts
by Khokhlov (2001) which predict the presence of unburned and partially burned material down
to the central regions, and a large amount of unburned material at the outer layers.
 Mergers are beyond the scope of this paper but pure deflagration models
will be mentioned where appropriate.

 The DD-model assumes that burning starts
 as subsonic deflagration and then turns to a supersonic, detonative mode of burning.
 The amount of $^{56}$Ni depends primarily on
$\rho_{tr}$ (H\"oflich 1995; H\"oflich; Khokhlov \& Wheeler 1995; Iwamoto et al. 1999),
 and to a much lesser extent on  the deflagration speed, and the
initial central density and initial chemical composition (ratio of
carbon to oxygen) of the WD.
 In DDs, almost the entire WD is burned, i.e. the total production of
nuclear energy is almost constant. This and the dominance of $\rho_{tr}$ for the $^{56}Ni$ production
are the basis of why, to first approximation, SNe~Ia appear to be a
one-parameter family. The  observed $M(\Delta M_{15})$ can be well understood
as an opacity effect, namely the dropping opacity at low temperatures
 (H\"oflich et al. 1996b and references therein; Mazzali et al. 2001).
 Nonetheless, variations of the other parameters  lead to some
deviation from a one-parameter $M(\Delta M_{15})$ relation with  a spread of $0.5 ^m$
(H\"oflich et al. 1996b).
 Empirically, the $M(\Delta M_{15})$  has been well established with a rather small statistical
error $\sigma$ ($0.12^m$: Riess et al. 1996;
 $0.16^m$: Schmidt et al. (1998); $0.14^m:$ Phillips 1999; $0.16^m:$ Riess et al. 1999;
$0.17^m:$ Perlmutter et al. 1999a).
  This may imply a correlation between free model parameters, namely
the  properties of the burning front, the main sequence mass of the progenitor
$M_{MS}$,  and   the central density of the WD at the time of the explosion.

 Recent studies have shown that the chemical structure of the WD will affect the 
LCs and spectra. The properties of the
progenitors must be taken into account to determine cosmological parameters or
 the cosmological equation of state. In particular, the mean C/O ratio of the exploding WD has been
identified as one of the key factors
(H\"oflich et al. 1998, 2000; Umeda et al. 2000; Dominguez, H\"oflich \& Straniero 2001).
 From stellar evolution, the WD can be expected to consist of a C-depleted central region produced
during the convective, central He burning. The central region is surrounded by layers with 
$C/O \approx 1$ originating from the He-shell burning in the star and from the accretion
phase. The size of the C-depleted region ranges from 0.1 to about 0.8 $M_\odot$ 
depending on  $M_{MS}$.  At the time of the explosion, the C/O-WD should show a distinct C and O profile, 
 These dependencies may suggest that the pre-conditioning of the WD may be key for our understanding of
the diversity of supernovae (H\"oflich et al. 1996). In addition, Arnett \& Livne (1994) suggested that
the initial velocity fields in the WD prior to the explosion might influence the flame propagation.

 The propagation of a detonation front is well understood
but the description of the deflagration front
and the deflagration to detonation transition (DDT) pose  problems.
 On a microscopic scale, a deflagration  propagates
due to heat conduction by electrons. Though the laminar flame speed in SNe~Ia is well known, the
front has been found to be Rayleigh-Taylor (RT) unstable, increasing
the effective speed of the burning front (Nomoto et al. 1976).
More recently, significant progress has been made toward  a better
understanding of the physics of flames. Starting from static WDs,
hydrodynamic calculations  of the deflagration fronts have been performed in 2-D
(Niemeyer \& Hillebrandt 1995, Reinecke et al. 1999, Lieswski et al. 2000) and 3-D
(e.g. Livne 1993, Khokhlov 1995, Khokhlov 2001).     It has been demonstrated that R-T instabilities
govern the morphology of the burning front in the regime of linear instabilities,
i.e. as long as  perturbations remain small.  During
the first second after the runaway, the increase  of the flame surface due to RT
remains small and  the effective burning  speed is close to the laminar speed ($\approx 50 km/sec$) if the
ignition occurs close to the center.
 Khokhlov (2001) also shows that 
 the effective burning speed is very sensitive to the energy release by
the fuel, i.e. the local C/O ratio.  Therefore, the actual flame propagation will depend
on the detailed chemical structure of the progenitor.
 Niemeyer, Hillebrandt \& Woosley (1996) studied the effect of off-center ignition and demonstrated that
 multiple-spot ignition with significant separation ($\approx 50 ... 100 km$) will significantly alter the early propagation of the flame.
 For strong off-center ignitions, the buildup time of RT-instabilities is shorter corresponding to the
larger gravitational acceleration. Still, even for fast rising blobs, their morphology and, consequently,
the effective burning speed will depend critically on small scale motions of the background (see below).

 Despite these advances, the mechanism is not well understood
 which leads to a DDT or, alternatively, to  a fast deflagration in the non-linear regime 
 of instabilities.
Possible candidates for the mechanism are, among others, the Zeldovich mechanism, i.e.
mixing of burned and unburned material (Khokhlov, Oran \& Wheeler 1997ab, Niemeyer \& Woosley 1997),
crossing shock waves produced in the highly turbulent medium, or shear flows
of rising bubbles at low densities (Livne, 1998). Currently,  none of the proposed 
mechanisms have been shown to work in the environment of SNe~Ia.
The Zeldovich mechanism leads  to a DDT only if the density and temperature fluctuations remain small
(Niemeyer  1999), and the effectiveness of crossing shock waves or shear flows has
yet to be demonstrated. However, as a common factor, all these mechanism 
will depends on on the  physical conditions prior to the DDT.
  
 From the analysis of LCs and spectra, and the study of flame fronts in SNe~Ia,
there are strong indications for the importance of the initial structure of the WD prior to the
nuclear runaway.

 In this work we present the evolution of the WD just prior to the thermonuclear runaway based on
multidimensional calculations. In particular, we want to address the following questions:
1) Do mixing processes change the chemical structure of the WD prior to the explosion?
2) Does the thermonuclear runaway occur in multiple spots?
3) Does the thermonuclear runaway happen in a static white dwarf?
 In Section 2 we present the setup of our calculations.
 In Section 3, the results are presented.
 In the final, concluding section, we discuss the results in the context of the modeling of SNe~Ia,
the use of SNe~Ia as cosmological yard sticks, and address the limits  of our  study.

\section{Numerical Methods and Setup}

 The initial model has been constructed  from results for the  stellar progenitor evolution  based on
 the code FRANEC (Straniero et al. 1988,  Chieffi et al. 1989, Limogni et al. 2000).
 The subsequent accretion phase on  the WD has been followed
 up to the thermonuclear runaway by solving the standard equations for stellar evolution
in a Henyey scheme (H\"oflich et al. 2000).
 Nomoto's equation of state is used (Nomoto et al. 1982).
 For the energy transport, conduction (Itoh et al. 1983),
convection  in the mixing length theory, and radiation are taken into account. Radiative opacities for free-free and
bound-free transitions are treated in Kramer's approximation and by free electrons. A nuclear 
network of 35 species up to $^{24} Mg $ is taken into account based on the reaction network of
Thielemann, Nomoto \& Hashimoto (1996).

 For our study of multidimensional effects, we start from a WD model several hours before runaway.
The spherical model is re-mapped on a spherical grid with 191
 radial  and 31 angular ($\Theta$) zones within a cone with
$\Theta $ between $45^o$ and $135^o$.
 The radial resolution has been decreased by a factor of $\approx 10$  from inner to the outer layer to properly resolve
the central regions. Note that the effective Reynolds numbers are of the order of 30 in the convective region. Thus,
we cannot resolve the small scale, turbulent motion but our study is limited to the large scale, convective flows.
 The initial structure is relaxed on this grid assuming
a pure carbon/oxygen mixture.
To carry out multi-dimensional simulations of
the interior convection;  and because of the low Mach number of the
associated flows, a compressible implicit hydro code is required.
 For available explicit, compressible codes for thermonuclear burning,
the sound crossing time over a resolution element limits the time step, i.e. CFL condition which
imposes hopelessly short time steps.
Therefore, for the further evolution, we use an implicit, Eulerian 2-D hydrodynamical code (Stein, Barkat \& Wheeler 2001).
The perpendicular velocity is assumed to be zero [reflective] at all boundaries.
 The hydrodynamical equations are solved in a $2^{nd}$ order scheme including first order centrifugal and Coriolis forces.
Radiation transport effects  have been neglected.
 For the equation of state, we use
a relativistic Fermi-gas  with  Coulomb corrections, and radiation.
Nuclei are treated as an ideal, non-relativistic gas without crystallization.
 For the nuclear burning,  analytic expressions have been used for the production of nuclear energy 
(Rakavy \& Shaviv 1968, Barkat et al. 1990), and
calibrated by an $\alpha $-network and tested against the detailed network given above
(Thielemann et al. 1996). The time step has been limited by the flow of the material from zone to zone.
Typically, the maximum exchange of matter is limited to $ \approx 5 \%$.
 We use standard quadratic artificial viscosity,
and small second-order Lax throughout the star. Small first order
Lax is used in the inner 3 rows of cells, and in the outer cells.
 Small scale fluctuations were introduced to initialize the convection. Tests showed that the results do no depend
on the initial spectrum of the fluctuations.

\section {Results}

 The structure of the initial model of the C/O white dwarf is based on a star with 3 solar
masses at the main sequence and solar metallicity which, at the end of its evolution, has lost
all of its H and He-rich layers.
 By accretion, its core has been grown  close to the Chandrasekhar limit 
 (Dominguez et al. 1998, Dominguez \& H\"oflich 2001).
The temperature, density and chemical profile at about 1 day before runaway (in 1-D) have been
re-mapped to the 2-D grid.
 At this time, the central density of the WD is $2 \times 10^{9} g/cm^3$.
  The carbon concentration
in the inner layers with 0.348 $M_\odot $ is a result of the central helium burning during the stellar evolution. For the outer
layers, the C/O ratio is close to 1. At this time, the nuclear burning times scales are of the order of days,
 i.e. much longer than the hydrodynamical time scales ($\approx 1 sec $).
 Because the WD is almost isothermal, the entropy is increasing with radius. 
The initial structure is shown in Fig.\ref{ini_ref} (after relaxation).
In the reference model, the further evolution has been followed up all the way to the
thermonuclear runaway.
 The computational domain extends  between 65 km to 2000 km in the radial direction.
  For the detailed study of the runaway, we use an extended computational domain down to 13.7 km.

\subsection{Evolution of the  Reference Model}

\subsubsection{The Mixing Phase}
 
 The nuclear time scales are long compared to the hydrodynamical
time scales up to a few seconds before runaway.
Because most of the mixing of abundances happens during this stage of long nuclear time scales
 (see below),
 it will be referred to as  'mixing phase'.

 In Figs. \ref{c_ref}, \ref{nuc_ref} \& \ref{e_ref}, snapshots of the evolution are shown for various quantities at  2 and 1 hours, 15 and 5 minutes
before the runaway.
 Nuclear burning of $^{12}C$ in the central region increase the entropy and temperature. Consequently, the rate
of energy production grows with time. The increase of
temperature by about 24 \% in the central regions
results in a decrease of the central density by about 4 \%.
  Most  of the nuclear energy contributes directly to an increase of the entropy .
In addition, the nuclear burning drives large scale, convective flows.
 Typical velocities of the convective flows increase from about 10 km/sec at 2 hours before runaway (b.r.)
  to $\approx 50 km/sec$ at about 5 minutes
b.r. (Fig. \ref{e_ref} and see below).
 The size of the eddies is comparable to the pressure scale height ($\approx 100 km$)
 (see Fig. \ref{c_ref}). The life time of individual eddies is of the order of one revolution for a mass element,
leaving little chance to produce a pattern typical for stationary convective layers.
 The flow is statistically steady on time scales short compared to the nuclear evolution time scales up
 to about 1 hour before the runaway.
   Eventually, most of the kinetic energy dissipates and contributes to heating.
 At the time of the explosion, the kinetic energy is small compared 
 to the nuclear energy produced ($7.01 \time 10^{45}$ vs. $  2.456\time 10^{48}$ erg).
 Initially,  the convective region is confined
by the  chemical boundary.
Later on, the entropy grows inside by nuclear burning, and a core of almost constant entropy
forms. The convection is confined by the steep entropy gradient as a consequence of the
steep entropy rise in  the outer layers (caused by the flat temperature profile in combination
with the rapidly decreasing density, see above).
 Due to the increase of the entropy with time,  this boundary is gradually moving
outwards in both mass and radius. Consequently,
 material of the carbon rich mantel and the core are mixed.
 The carbon concentration in the center increases from 24.7 to 35.6 \% at the time of the runaway and  its size
grows from 540 to 730 km (Fig. \ref{c_time_ref}). Due to the small energy production, hardly any mixing occurs early on but
the rate of mixing and of the change of R, i.e. the contour with C=45\%,
 grows strongly with time as the nuclear burning increases.
 No evidence is found for rising blobs which pass the sharp boundary of convective and non-convective layers
 and stay there. Blobs which pass the boundary tear pieces from the non-convective layer, creating a
new sharp boundary.
In stellar evolution,  penetration of individual elements of about 0.2 to 0.25 pressure scale heights is assumed
by some authors (e.g. Bressan et al. 1993, Schaller et al. 1992).
 If present, this effect  would result in a chemical
mixing of the entire WD with time scale of a few hours.
 In our example, a significant fraction of the WD mass is
enclosed within about 3 scale heights above the chemical inhomogeneity which corresponds to
a distance of 500 km. If we assume h=0.25 and a velocity of 10 km/sec for the turbulent eddies,
the corresponding time scale for complete mixing would be $\approx 3 h$.

 For any given progenitor structure, the lack of passing blobs through the boundary of convective and non-
convective region allows us  to estimate the total amount of mixing even without detailed calculations.
 During the phase of slow burning, only a negligible fraction of the C is consumed,
 the turbulent region is confined by the  steep entropy gradient at the chemical 
 boundary, and the entropy is almost constant within the turbulent center
(Fig. \ref{ce_ref}). Nuclear runaway occurs in our model when the mean entropy in the core
increases to $\approx 10.4 $. Nuclear burning increases the level of entropy in the core.
 We can estimate the final amount of mixing and the  radius of
 the core
by identifying the distance in the initial model at which the entropy correspond to the
critical value for the runaway. Our estimate hardly depends on the exact value
of the entropy at the runaway due to the steepness of the entropy gradients.

\subsubsection{The Nuclear Runaway: The Last 5 Minutes before Runaway}

In Fig. \ref{t_ref}, we show the final evolution of the temperature and the velocities.
Increasing nuclear burning in the inner 100 km drives increasingly strong convection.
The region of enhanced energy production
heats up material. This hot material starts to rise.
 Typical turbulent velocities  increase  from $\approx 50$ to $\approx 100 km/sec $
at the time of the runaway. 

 The unsteady convective flows are a key factor to understand the trigger for the final thermodynamical runaway.
  It explains why we do not see ignition in multiple spots or a strong `off-center' ignition.
Due to convective mixing, the entropy remains nearly constant in all but the
very inner layers with a central distance $\lesssim 150 km$.  There,
large scale, convective motion  brings in material radially. Eventually,  compressional heat
causes the thermonuclear runaway close to the minimum distance of the corresponding eddy.
 In the reference model, the thermonuclear runaway occurs very close to the inner boundaries
at a distance of 65km in one specific cell. 
In general, we do not expect multiple spot ignition 
because  the size of the eddies is larger than the central distance of the point where the runaway occurs.
We note that the convection also operates at larger distances from the center but, there,
the relative changes in radius and, therefore, the release of  compressional heat
is insufficient to bring material to the point of explosive nuclear burning.
 We expect that the thermonuclear runaway occurs earlier than in 1-D models in which
it is triggered by the  overall compression of the WD.

 To study the thermonuclear runaway in more detail, we have recalculated the final stages up
to the runaway for the same model, but the computational region has been extended down to 13.7 km.
 For computational efficiency,  we started with
an increased rate of nuclear reactions by  factors  of 200 down to 4
up to about 10 minutes and 127 seconds before the explosion, respectively.
 At 127 seconds, the resulting structure resembles very closely the reference model
  for the entropy, the density and the chemical structure because
 the main effect of the nuclear burning is an increase of the entropy,
the resulting mixing, and the short life time of convective cells (see sects. 3.1.1 and 3.1.3).
 However, the increased heating
 results in a slight increase of the total kinetic energy at t=127 sec by about 10 \%
 compared to the reference model.
 Fig. \ref{cte_high} shows the final structure at the onset  of the thermonuclear runaway for the inner 120 km.
Any deviations from a radial structure are limited to this inner region.
Eventually, the thermonuclear runaway occurs in one cell at about a distance of 27 km.
During this last phase,
the strong release of nuclear energy drives large scale of violent motion of the matter.
 The pattern in the temperature
and energy release follows the large scale motion. Most noticeable is the C-like pattern 
 of the temperature distribution close to the
center (Fig. \ref{cte_high}, lower left panel). The density shows only very minor deviations from sphericity.
 At this stage of evolution, Carbon is locally depleted  by about 1 to 2 \% due
to the nuclear burning and, again, it is carried by the velocity field.

We want to discuss  the evolution to the thermonuclear
runaway in some detail. As shown above, the structure  of the entropy, temperature, nuclear energy production
and chemistry can by understood in the same way as a result of the convective motion.
 As an example, the evolution of the temperature and the velocity field is shown in Fig. \ref{t_high}.
In the following, the coordinates in brackets provide the coordinates of features
in the (x,y) plane in km.
 At about 3.54 seconds (fig. \ref{t_high}, left upper plot) before the runaway, the temperature structure starts to deviate from the
radial structure. The high velocity field at the upper part of the plot (red arrows) is part of a
larger vortex A  with the center at  (150km,-10km) which extends down to about 70 km. Close to the
center of the WD,
nuclear burning drives a convective flow in the opposite direction. These two regions are separated
by a layer with higher temperature and low velocities. Due to the shear, a new, small vortex B is
evolving at (-15km,+65km). It results in a redirection of the material flow at the lower 
edge of vortex A.
 This material flow is directed inwards, and compresses and heats up
 material in front of the flow pattern.
 At about 0.353 seconds, the temperature has risen up to $8.5\times 10^8 K$. Eventually,
further compression and burning causes a rise in temperature up to the onset of explosive carbon burning
($ \approx T \geq 1.3 \times 10^{9} K$).

   Previously, Garcia-Senz \& Woosley (1994) studied the details of the thermonuclear runaway in 1-D. They considered
plumes rising in a static background. They found that the runaway occurs in rising plumes which rise with velocities
of about 5 to 30 km/sec at central distances of 30 and 100 km/sec, respectively.
 The runaway occurred when the increase in the thermonuclear burning 
in the plume becomes stronger than the cooling by expansion. In our simulations, we similarly see that
plumes with increased burning tend to rise close to the thermonuclear runaway. In our calculations,
these plumes form close to the central region ($\leq 30 ... 60 km$) due to the temperature increase close to the center.
 At times close to the runaway, the nuclear energy production in the plumes almost compensates for the cooling.
However, in a moving background (with  velocities $\leq 100 km/sec$), the rising plume will be disrupted and parts find themselves
 in both a rising and descending velocity fields.
  For those parts that go downward by the current, adiabatic expansion
will not avoid the runaway but, in contrast, compression will push the element over the 'edge'.
 In a non-stationary WD, the thermonuclear runaway will occur slightly earlier than in a static WD.

\subsubsection{Effects of the Nuclear Reaction Rates}

 We have studied
 the sensitivity of our results to the assumptions and uncertainties related to
 the nuclear energy production. In particular, the reaction  $^{12}C(\alpha, \gamma )^{16}O$  
 must be regarded as
 uncertain by a  factor of three (e.g. Buchmann, 1997) despite some indirect evidence which
favors a large cross-section. This indirect evidence stems from
recent studies  of pulsating WDs (Metcalfe, Winget \& Nather 2001) and 
 from the rise times of light curves of SNe~Ia (H\"oflich et al. 1998, Dominguez et al. 2001).
Stellar evolution for the asymptotic giant branch
 favors also a high cross section for  $^{12}C(\alpha, \gamma )^{16}O$,  but a low
value can be compensated for by an increased mixing of helium into the stellar core
(e.g. Salaris et al. 1993).

 We have scaled the rate for the nuclear energy production by   factors $ f$ between 1 and 200
(Table \ref{table1}).
For the same initial model, a higher production of nuclear energy drives a  faster convection
(see $E_{kin}$ in Tab. 1),
and it decreases the time till the thermonuclear runaway, i.e. the nuclear
burning time scales are reduced by the factor $f$. This leaves less time for mixing coupled with
increased  fluctuations of the central carbon abundance, so a large $f$ implies
that a lower compression is required to trigger the explosion.
 In all calculations, single spot ignition has been found.  However,  the  central
distance of the thermonuclear explosion increases from 27 km $(f=1)$ to about 90 km $(f=200)$, and the 
typical, convective velocities
increase slightly.
The fluctuations in the carbon concentration rise from the 1 \% level to about 5 \%.
 The amount of carbon mixing decreases with an increasing reaction rate but, overall,
it is rather insensitive for $f \leq 20 $ (Fig. \ref{c_high}).

\subsection{ Final Discussions and Conclusions}

We have studied the final hours of a Chandrasekhar-mass WD prior to the
thermonuclear runaway to investigate the pre-conditioning of exploding WDs, namely
chemical mixing and the ignition process.
 
 The initial model has been constructed from results of stellar evolution for a star with 3 solar
masses with solar metallicity, followed by a subsequent accretion phase
close to the onset of the thermonuclear runaway (Dominguez et al.  2001).
 The WD has a mass of
$\approx 1.37 M_\odot$. Its chemical structure is characterized by a central region of $0.348 M_\odot$
with a low C-concentration ($\approx 24 \%$) surrounded by a mantel  with $C/O \approx 1$
originating from the He-shell burning and the phase of accretion. A few hours before runaway,
the thermal structure of the progenitor shows a rather flat temperature profile, and a steep
entropy profile because the rapidly dropping density. 
 
 Prior to the runaway, the central regions undergo mild C-burning. The resulting
energy release drives convective motion in the inner region of low C-concentration and, gradually,
increases the entropy of the core up to the point of ignition.
 Due to the convection, the entropy of the core is almost constant.
 Within the resolution of our models, the carbon-concentration  gradient at the boundary
between the core and the mantle prevents direct mixing,  e.g. due to overshooting convective
elements. However, the increasing entropy of the core results in a negative 
entropy gradient at the core boundary which compensates for the
carbon-concentration gradient.  This increases the
region with constant entropy and produces mixing of C-rich region into the
core with typical fluctuation of about 1\%.
 We find that the central C-abundance increases from 24 to about 37 \%. The initial WD
is not homogenized, but the jump in the carbon abundance is reduced by a factor of $\approx 2$.

 At the time of the explosion,
 a pattern of large scale, convective elements has been established
 with sizes of typically  100 km and convective velocities
between $\approx 40 $ to $\approx $ 120 km/sec.
 Differential velocities between adjoining eddies are larger by a factor of 2 which
is well in excess of the laminar deflagration speed.
 Thus, the change of the morphology of the burning
front of SNe~Ia is determined by the pre-conditioning of the
WD during the early phase of the explosion for $\approx$
 0.5 to 1 sec (Dominguez \& H\"oflich 2000, Khokhlov 2001).
 Niemeyer et al. (1996) found significantly shorter time scales for the growth of RT-instabilities
in their study of strongly off-center explosions. This can be expected as a result of the larger gravitational
acceleration.
 Still, even for their fast rising, large scale  blobs ($\approx 1000 km/sec$), the morphology of the plumes
 and, consequently, the effective burning speed will depend critically on small scale motions of the background.
 The effective surface of the front will be increased resulting 
in significantly higher burning speeds.
 Faster burning implies a larger region of low
proton to nucleon ratio and, thus, a larger production
of neutron rich isotopes in the central region. On the other hand,
 a reduction of the time scales for electron
capture can be expected  leading to
the production of less neutron rich isotopes.
 Possible consequences for current estimates on
 the limits  on the central densities of the
WD should be noted (Brachwitz et al. 2000).

 The explosive nuclear burning front starts in one well defined
region close to the center ($\approx 30 km$).
 The size of the ignition region is determined by the
grid resolution ($\approx  2 km$).
The explosive phase of burning is triggered by compressional heat 
when matter is brought inwards by convection.
 It starts close to the center because, there, the adiabatic heating combined with thermonuclear reactions
are most   effective for a given size of turbulent elements.
 We find no evidence for multiple spot or strong off-center
ignition.  We do not expect it because the size of the eddies is comparable to the
central distance of the ignition point, and
the lack of any mechanism which would cause a synchronization
within typical time scales for the runaway ($\leq 0.1 sec$). Thus, the
 probability is fairly small for  having a second ignition point during that time.

 In the following, we want to put our basic results into context for our understanding
and the quantitative modeling of SNe~Ia.
As mentioned in the introduction, the propagation of the deflagration front depends on 
the energy release and, consequently on the fuel (Khokhlov 2001). We find that
the chemical profile in the WD will be strongly changed, but in a predictable way.
 We find that the initial velocity field must be expected to alter the flame propagation
during the deflagration phase. 
 Although the actual deflagration speed has little effect on the overall chemical structure
of DD-models with the exception of the production of neutron rich isotopes close to the center,
 all proposed mechanisms
for the DDT identify the pre-conditioning of the material during the deflagration phase as a
key element (see introduction) which, in turn, is strongly effected by the initial WD.
 
 We may suspect from the comparison between normal bright and subluminous SNe~Ia and
the role of the DDT transition for the brightness decline relation that the precondition
of the WD may be the `smoking gun' for our understanding of the diversity of SNe~Ia.

 As mentioned above, the overall chemical structure of the initial WD is preserved, and the
turbulent velocity field is limited to the inner, C-depleted core. Both
the velocity field and the C-concentration influence the  burning front.
Therefore, the mass of the
progenitor has a direct influence on the outcome because the core size depends mainly on the
$M_{MS}$ mass  of the progenitor.
 The consequences are obvious with respect to the evolution of the  SNe~Ia with redshift 
 and their use as a yardstick
to measure cosmological parameters and the cosmological equation of state.

Finally, we have also to mention the limitations. This study should not be seen as a final answer
but as a starting point to open a new path which, eventually, may lead to a deeper understanding
of the relation between the progenitor and the final thermonuclear explosion.

This work is based on hydrodynamical simulations in two rather than three dimensions.
In either case, the convection is driven by entropy gradients over large distances.
These large gradients drive large eddies. Convective eddies
lose energy both in the true 3-D case  and in our 2-D simulations  mostly by exchange with 
eddies of different size but the mechanism and rate of energy loss differ between 2-D and 3-D (see below).
 Interaction between eddies of different sizes causes exchange
of energy towards larger and smaller eddies (Porter \& Woodward 1994).

It is well known that, for a fully developed turbulence in an incompressible fluid,
the direction of the {\it average} energy flow is from large scale eddies
to smaller ones in 3-D, and from small scale eddies to larger ones in 2-D  (Kraichnan 1967, 
Rose \& Sulem 1978, Kraichnan \& Montgomery 1980).
 The viscosity of the fluid becomes most important and, thus, the  dissipation of kinetic energy is most efficient for
the smallest eddies whose Reynolds number is comparably with unity
 (Laundau \& Lifshitz 1989).  For incompressible fluids, the dissipation of kinetic energy is very different in 2-D and 3-D.
In the limiting case of vanishing viscosity and incompressible fluids, the dissipation rate in 3-D remains finite
while it approaches zero in 2-D because the inverse cascade in the energy flow.
 The different behavior of the {\it average} energy flow in 2-D and 3-D  is caused by quadratic invariants 
globally conserved by the advection term  in the hydro equations (see e.g. Hasegawa 1985 and, more general,  Vazquez-Semadeni 1991).
These quantities are  {\sl not} conserved in compressible fluids where the difference between 2-D and 3-D will be of a
different kind. Namely, energy can dissipate by acoustic waves and shocks, in addition to the energy dissipation  by
viscosity.
  Thus, the interaction of the largest eddies in a finite space and in case of a compressible fluid is less clear.

 In our simulations, the decay time scales of large eddies are of the order of one rotation.
  The real viscosity of  astrophysical fluids is much smaller than the numerical viscosity in our simulations.
At the same time, large scale flows (rolls) have a higher inertia in 2-D than in 3-D.
We do not know whether the lifetimes of the true large 3-D eddies are larger or smaller
than in our simulations but we argue that the decay times of one rotation may be the right
order of magnitude. In the following paragraphs, we argue that the main results hold.
 No detailed simulations for 3-D are available for  the conditions 
in Chandrasekhar mass WDs and  subsonic convection.
 However, 3-D studies and simulations for convection
in other environments suggest  dissipation time scales very similar to our results.
 For incompressible fluids, the  dimensional analysis suggests decay time scales of the order
of the revolution time  in the largest eddies (Landau \& Lifshitz, 1989).
 For fully compressible thermal convection in deep atmospheres, Porter \& Woodward (2000)
extended their 2-D to 3-D, and found similar results in both cases.
 Recent simulations for the supersonic case and MHD turbulence
 by Stone, Ostriker  \& Gammie (1998) indicate typical
time scales for the energy dissipation in molecular clouds of about 0.3 to 0.8 revolutions 
for large eddies, i.e. about 1.5 times faster compared to similar, 2-D
calculations by Ostriker, Gammie \& Stone (1999).

 A further restriction is our resolution which  is limited to  Reynolds numbers of
$\approx 30$ to $50$, i.e. not sufficient to follow the cascades to small scales.
 Obviously, a high resolution, full three dimensional study would be desirable.
 In spite of this, we expect no qualitative change of our basic conclusions concerning 
 the mixing and ignition process.

  It is well known that the mixing properties of 2-D and 3-D unsteady flows differ, both 
  for scalar and vector "contaminants".
 In 2-D, each large eddy is a huge torus carrying a mass which is a large portion of the convection zone, and
one or two large eddies carry the hot material from the burning
center and spread it over the convective zone.
In 3-D, each large eddy  carries much less mass. Nevertheless, even in 3-D,
the number of large eddies should be sufficient for spreading material over the entire convective zone
because mixing continues for hours compared to the few seconds it takes a mass element to cross the
convection zone.
 As mentioned in the last section, the amount of mixing can be understood in terms of the 
 nuclear burning  which  increases the entropy in the central region both in 2-D and 3-D
calculations and, therefore, we expect a similar amount of mixing in both cases despite the differences
in the mixing properties.  Though, one may expect some change in the size of the fluctuations in 
the C-abundance and entropy (see above).

 We do not expect a qualitative change in size of the large eddies  at
the time of runaway and, therefore, in the ignition process, because the presence  of large eddies
is determined by their production.
The motion is continuously driven by nuclear burning at the innermost layers which produces a rise of
heated material over about a pressure scale height. This determines the size of the largest eddies which
must be expected to be of similar size in both 2-D and 3-D.
Close to the thermonuclear runaway, the circulation times become larger than the nuclear
burning time scales. Unless, in 3-D, the decay times of large eddies
are much shorter than a revolution, the largest eddies must be expected
to trigger the ignition in a way similar to the 2-D case.
 Therefore, the probability of ignition in more than one well defined  region remains small.
 Due to the resolution of our simulation, this region
has a size of several kilometers. We cannot say anything about the ignition process on scales 
of the nuclear burning front.

 We have discussed possible implications for the deflagration front on SNe~Ia based on
previous studies. Obviously, there is a need for consistent calculations of the deflagration front
to quantify our estimates on the propagation of the nuclear flames.

 Our results are based on a specific progenitor
 with a main sequence mass of $3 ~M_\odot$.  Similar studies may be useful
for other $M_{MS}$ with larger or smaller cores with low carbon abundances, and
different central densities at the time of the explosion.
 In light of the analysis of the
subluminous SN1999by (Howell et al. 2001), other effects such as rotation or crystallization
 should be considered in the future.

 \subsection*{ACKNOWLEDGMENTS}
We would like to thank Z. Barkat, A. Glasner,  A. Khokhlov, E. Livne, J. Scalo and J.C. Wheeler for
helpful discussions and useful comments, and R.  Hix for providing the calibration of the
analytic energy production rates. J. Stein would like to thank J.C. Wheeler and
the colleagues at the  Department for Astronomy for the hospitality 
during his sabbatical in 1995-6 when this work was started, and  during subsequent visits in 1997-2000.  P.H. would like to thank the
colleagues at the Racah Institute in Jerusalem for the hospitality during his visit.
This research was supported in part by  NASA Grant LSTA-98-022, and the John W. Cox Endowment Fund
to the Department of Astronomy at  UT.

\newpage

\begin{table}[t]
\begin{center}
\caption{Influence of the enhancement factor $F$ of the nuclear reactions on
the distance r(ignition) at which the thermonuclear runaway occurs,
on the mean carbon-concentration $C$,  and the kinetic energy at the onset
of the runaway. 
}
\begin{tabular}{llll}
\hline                     
Factor $F$  &          $  C$ &       r(ignition) & $E_{kin}$ \\
\hline                     
\+ 200  &     26+-1  \%    &      90 km     &  -      \\
\+ 50    &   32.5   \%   &      86 km      &   -  \\
\+  20    &   35.5 \%   &        71 km    &  8.5E+45 erg      \\
\+   4     &  36.2 \%  &         32 km    &  8.2E45 erg  \\
\+   1      & 37.0  \%  &          27 km  &   6.9E45 erg  \\
\hline                     
\end{tabular}
\end{center}
\label{table1}
\end{table}

\newpage
\begin{figure}
 \caption{
 Density,  carbon concentration, nuclear reaction rate and entropy (in CGS)
are given for the WD with a radius of 1800 km at 3 hours before the runaway for the 2-D model,  i.e.
about  15 minutes after the start of the 2-D calculations. Up to this time, the changes
in the chemical structures are negligible. The coordinates are given in {\sl cm} relative to
the center of the WD.
 The computational domain in radius and angle
is  65 to 2000 km and $45^o$ to $135^o$, respectively. The horizontal axis is the axis of symmetry.
 Note that the carbon abundance in the outer layers is 0.5, i.e. outside the color range.
}
\label{ini_ref}
\end{figure}
\begin{figure}
\caption{
Evolution of the structure during the `mixing' phase.
  Carbon concentration at the inner layers of the WD as a function of time before runaway (b.r.).
 The coordinates are given in cm relative to
the center of the WD. }
\label{c_ref}                 
\end{figure}
\begin{figure}
\caption{
 Same as Fig. \ref{c_ref}, but for the nuclear reaction rates.
}
\label{nuc_ref}
\end{figure}
\begin{figure}
\caption{
 Same as Fig. \ref{c_ref}, but for the entropy. In addition, the velocity fields are given.
Black, red and green vectors correspond to velocities in the ranges between  0 to 20 km/sec,
20 to 40 km/sec, and 40 to 60 km/sec, respectively.
}
\label{e_ref}
\end{figure}
\begin{figure}
\caption{
Mass fraction and  size of the region with low C ($X(C) \leq 45 \%$) as a function of time.
}
\label{c_time_ref}
\end{figure}
\begin{figure}
\caption{
 Carbon concentration (left)  and entropy (right) at three hours (upper)
  and one minute (lower) before
runaway. Runaway occurs when the mean entropy in the turbulent core rises to about 10.1.
Note that the final size of the mixed region corresponds to the distance at with the
same entropy can be found in the initial model (see text).
}
\label{ce_ref}                
\end{figure}
\begin{figure}
\caption{
Final evolution of the temperature structure up to the runaway.
 In the lower left plot, the runaway occurs in the second red zone
  from the left right at the  inner boundary  
  (T=$1.74 \time 10^{9}$ K).
  In addition, the velocity field  is  given.
Black, red and green vectors correspond to velocity ranges of  0 to 50,
50 to 100, and 100 to 150  km/s, respectively.
}
\label{t_ref}                 
\end{figure}
\begin{figure}
\caption{
Final evolution of the temperature structure up to the runaway for a model with the same 
physics as the reference model but with a smaller inner `core' of 13.7 km instead of 65 km.
 For computational efficiency, the nuclear rates have been increased up to about 1 minute before
the runaway. The runaway occurs in the left, upper red cell (T=5.14E9 K). The cell has a size of
$\approx 2 km$ and it is  at a central distance
 of 27km. The neighbouring red cell has a temperature of 1.06E9 K, i.e. before runaway.
  In addition, the velocity field  is  given.
The longest vectors in black, red and green correspond to velocities of  50,
 100, and  150  km/s, respectively. Typically, the velocities are between 30 and 60 km/sec.
}
\label{t_high}                
\end{figure}
\begin{figure}
\caption{
Same as Fig. 8, but
carbon concentration, temperature, nuclear energy generation and entropy at the
 runaway for the model with an inner cavity of 13.7 km. In this cell
 C (and O) has been depleted by the explosive nuclear burning.
}
\label{cte_high}              
\end{figure}
\begin{figure}
\caption{
Effect of the nuclear reaction rate on C-abundance  and the distance at which
the runaway occurs for a nuclear reaction rate increased by a factor of 4 (upper) compared to
1 (lower).
}
\vskip  10.cm
\label{c_high}                
\end{figure}
\end{document}